\documentclass[a4paper,fleqn]{cas-sc}

\usepackage[sort,compress,numbers]{natbib}

\usepackage{hyperref}
\hypersetup{
  colorlinks = true,
  citecolor=blue,
  linkcolor=blue,
  urlcolor=blue
}

\usepackage{float}

\usepackage{bm}

\usepackage{xcolor}

\graphicspath{{fig/}{./fig/}{.}}

\begin{document}
\let\WriteBookmarks\relax
\def\floatpagepagefraction{1}
\def\textpagefraction{.001}

\shorttitle{Electronic properties and surface states of RbNi$_2$Se$_2$}    

\shortauthors{S. Basak et~al.}  

\title [mode = title]{Electronic properties and surface states of RbNi$_2$Se$_2$}  

\author[1]{Surajit Basak}[orcid=0000-0002-0669-0984]

\ead{surajit.basak@ifj.edu.pl}

\credit{Validation, Formal analysis, Investigation, Writing - Review and Editing}

\address[1]{
Institute of Nuclear Physics, Polish Academy of Sciences, 
ul. W. E. Radzikowskiego 152, 31-342 Kraków, Poland
}

\author[1]{Przemys\l{}aw Piekarz}[orcid=0000-0001-6339-2986]

\ead{piekarz@wolf.ifj.edu.pl}

\credit{Validation, Formal analysis, Investigation, Writing - Review and Editing}

\author[1]{Andrzej Ptok}[orcid=0000-0002-5566-2656]
\cormark[1]
\ead{aptok@mmj.pl}
\ead[URL]{www.mmj.pl/~aptok/}

\credit{Conceptualization, Methodology, Validation, Formal analysis, Investigation, Writing - Original Draft, Writing - Review and Editing, Visualization, Supervision}

\cortext[cor1]{Corresponding author}

\begin{abstract}
Iron-based superconductors, with the ThCr$_{2}$Si$_{2}$-type tetragonal structure (122 family), due to the iron arsenide/selenide layers exhibit several characteristic electronic properties.
For example, multiband character mosty associated with the {\it d}-orbitals of iron and the quasi-two-dimensional (2D) cylindrical Fermi surface. 
Moreover, external hydrostatic pressure leads to the isostructural phase transition from the tetragonal to collapsed-tetragonal phase.
In this paper, in relation to the iron-based superconductors, we discuss the electronic properties of novel 122-family member RbNi$_{2}$Se$_{2}$ [Liu H. {\it et al.}, \href{https://doi.org/10.1103/PhysRevB.106.094511}{Phys. Rev. B {\bf 106}, 094511 (2022)}].
We show that the two Fermi pockets exhibit quasi-2D character.
Calculation of the surface spectral function for the (001) surface shows that the surface states are realized independently on the surface termination.
Additionally, contrary to the iron-based 122 compounds, RbNi$_{2}$Se$_{2}$ exhibits extraordinary multiple isostructural phase transitions under pressure.
Moreover, the Lifshizt transition occurs under external pressure, which results into a strong modification of the shapes of the Fermi pockets.
\end{abstract}


\maketitle

\section{Introduction}

Discovered in 2008, the superconductors based on iron~\cite{kamihara.watanabe.08} opened a period of intensive experimental and theoretical studies of this class of compounds~\cite{ren.zhao.09,stewart.11,kordyuk.12,hosono.kuroki.15,fernandes.chubukov.16}.
These materials are characterized by the layered structure containing iron-arsenide/selenide layers -- where the iron square lattice is decorated by anions.
As a consequence, the electronic structure exhibits multiband properties and the characteristic quasi-two-dimensional (2D) cylindrical Fermi surface.

One of the examples of iron-based superconductors is a group of the ternary ThCr$_{2}$Si$_{2}$-type compounds (called 122 family)~\cite{luo.wang.08,neupane.richard.11,rotter.tegel.08,rotter.tegel.08b,tegel.marianne.08,shirage.miyazawa.08,anupam.paulose.09,alireza.alireza.ko.08,huang.qiu.08,leithejasper.schnelle.08,harrison.mcdonald.09,baek.lee.09,shirage.kihou.09,kawashima.ishida.18}.
The 122-family members exhibit not only superconductivity, but also a wide range of physical properties, e.g., antiferromagnetism~\cite{huang.qiu.08,leithejasper.schnelle.08,baek.lee.09,harrison.mcdonald.09}, spin density waves~\cite{luo.wang.08,neupane.richard.11,rotter.tegel.08}, and structural phase transitions~\cite{tegel.marianne.08}.
Moreover, in the presence of external hydrostatic pressure, the isostructural transition from the tetragonal to collapsed-tetragonal phase occurs~\cite{kreyssig.green.08,uhoya.stemshorn.10,mittal.mishra.11,uhoya.tsoi.10,tafti.juneaufecteau.13,taufour.foroozani.14,tafti.clancy.14,tafti.ouellet.15,wang.matsubayashi.16,ptok.sternik.19,ptok.kapcia.20,tresca.profeta.17}.

Recently, the RhNi$_{2}$Se$_{2}$ compound, showing the multi-gap superconducting properties below $T_{c} =1.2$~K, was successfully synthesized~\cite{liu.hu.22}. 
Contrary to a large part of the iron-based materials~\cite{gurevich.11}, RhNi$_{2}$Se$_{2}$ exhibits the Pauli paramagnetic behavior.
The initial measurements of the specific heat suggest enhancement of the effective electronic mass $m^{\ast} = 6 m_{e}$ and a weak role of correlations.

In this paper, on the base of density functional theory (DFT) calculations, we discuss the electronic properties of RbNi$_{2}$Se$_{2}$ (the band structure and the Fermi surface).
From the DFT band structure, we construct the tight banding model in the basis of the Wannier orbitals and investigate the surface states.
Mainly we discuss the surface termination dependence of the surface states.
Finally, we demonstrate the existence of the multiple-isostructural phase transitions occuring under external hydrostatic pressure.
The presented results can also be applicable for other 122-compounds based on nickel, e.g. KNi$_{2}$Se$_{2}$~\cite{neilson.llobet.12}, CsNi$_{2}$Se$_{2}$~\cite{chen.yang.16}, or TlNi$_{2}$Se$_{2}$~\cite{wang.dong.13}.

The paper is organized as follows.
The computational details are present in Sec~\ref{sec.method_theo}.
Next, in Sec.~\ref{sec.results} we present and discuss our results:
electronic properties (Sec.~\ref{fig.band}), realized surface states (Sec.~\ref{sec.surf}), and properties under hydrostatic pressure (Sec.~\ref{sec.press}).
Finally, a summary is provided in Sect.~\ref{sec.summary}.

\section{Computational details}
\label{sec.method_theo}

The electronic properties were calculated using
{\sc Quantum ESPRESSO}~\cite{giannozzi.baroni.09,giannozzi.andreussi.17,giannozzi.baseggio.20}.
The calculations were performed within the generalized gradient approximation (GGA) in the Perdew, Burke, and Ernzerhof (PBE) parameterization~\cite{pardew.burke.96} within {\sc PsLibrary}~\cite{dalcolrso.14}.
In this study we used the experimental lattice parameters~\cite{liu.hu.22}.

The DFT results of the electronic band structure calculation, performed for the primitive unit cell, were used to find the tight binding model in the basis of the maximally localized Wannier orbitals~\cite{marzari.mostofi.12,marzari.vanderbilt.97,souza.marzari.01}. 
It was performed using the {\sc Wannier90} software~\cite{mostofi.yates.08,mostofi.yates.14,pizzi.vitale.20}.
In our calculations, we used the $6 \times 6 \times 6$ full ${\bm k}$-point mesh, starting from the $p$ orbitals of Se, and from the $s$ and $d$ orbitals for Rb and Ni, respectively.
Finally, the resulting $28$-orbital tight binding model of RbNi$_{2}$Se$_{2}$ was applied to investigate the surface Green's function for a semi-infinite system~\cite{sancho.sancho.85}, using {\sc WannierTools}~\cite{wu.zhang.18} software.

Properties of the system under external hydrostatic pressure were investigated using the projector augmented-wave (PAW) potentials~\cite{blochl.94} implemented in the Vienna Ab initio Simulation Package ({\sc Vasp}) code~\cite{kresse.hafner.94,kresse.furthmuller.96,kresse.joubert.99} within the GGA PBE.
The energy cutoff for the plane-wave expansion was set to $400$~eV.
Optimizations of structural parameters (lattice constants and atomic positions) under pressure were performed in the conventional unit cell (containing two formula units) using the $12 \times 12 \times 4$ {\bf k}--point grid in the Monkhorst--Pack scheme~\cite{monkhorst.pack.76}.
The convergence criteria were set to be $10^{-6}$~eV and $10^{-8}$~eV for ionic and electronic degrees of freedom, respectively.

\section{Results and discussion}
\label{sec.results}

RbNi$_{2}$Se$_{2}$ crystallizes in the ThCr$_{2}$Si$_{2}$-type structure (symmetry I4/mmm, space group No.~139)~\cite{liu.hu.22}.
In the absence of any external pressure, the lattice parameters were experimentally measured to be $a = b = 3.9272$~\AA, and $c = 13.8650$.
In this cell three nonequivalent atoms Rb, Ni, and Se are placed at the crystallographic sites: $2a(0,0,0)$, $4d(0,1/2,1/4)$ and $4e(0.0,z_\text{Se})$, respectively.
Experimentally, the free parameter at the position of Se atom was determined to be $z_\text{Se} = 0.3498$.
The lattice parameters are comparable to the parameters of the isostructural KFe$_{2}$As$_{2}$ superconductor~\cite{ptok.sternik.19}.

\begin{figure}[!h]
\centering
\includegraphics[width=0.6\linewidth]{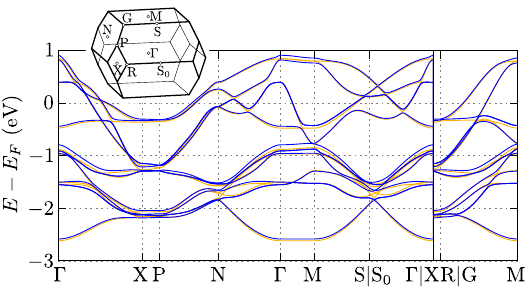}
\caption{The electronic band structure of RbNi$_{2}$Se$_{2}$. The results in the absence and presence of spin--orbit coupling are represented by orange and blue lines, respectively. Inset presents the Brillouin zone and its high symmetry points.}
\label{fig.band}
\end{figure}

\subsection{Band structure and Fermi surface}
\label{sec.band}

\begin{figure}[!t]
\centering
\includegraphics[width=0.5\linewidth]{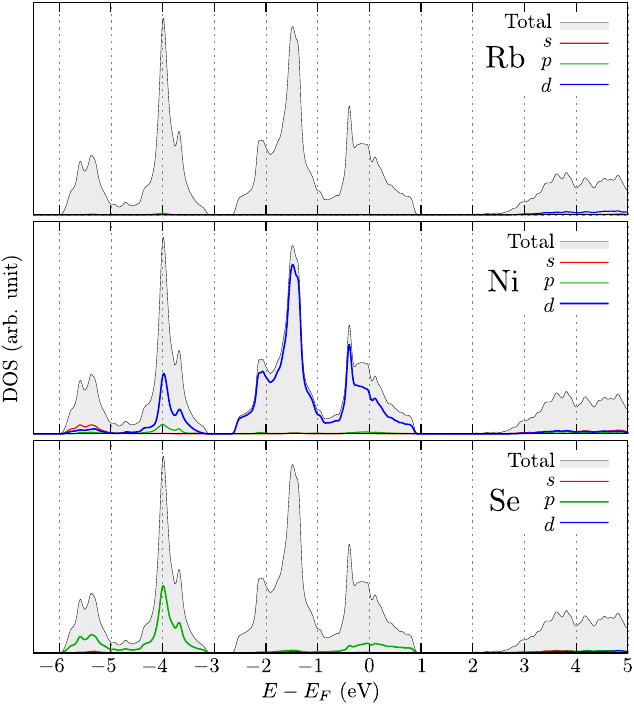}
\caption{
The total and partial electronic density of states of RbNi$_{2}$Se$_{2}$.
}
\label{fig.dos}
\end{figure}

The electronic band structure (in the absence and presence of the spin--orbit coupling) is presented in Fig.~\ref{fig.band}.
In relation to the 122 iron-based compounds, e.g. KFe$_{2}$As$_{2}$~\cite{ptok.sternik.19}, the Fermi level is shifted to the higher energies.
This is associated with the three ``extra'' electrons in the system (due to the Fe$\rightarrow$Ni and As$\rightarrow$Se substitutions).
Nevertheless, the Fermi level is still located within the range of energies related to the {\it d}-orbitals~\cite{liu.hu.22}.
This behaviour is also well visible in the partial density of states, presented in Fig.~\ref{fig.dos}.
For energies from $-6$ to $1$~eV, the Se {\it p}-orbitals and Ni {\it d}-orbitals contribute the most.
The group of the {\it d}-orbitals are mostly locallized between $-3$ and $1$~eV, and constitute a group of bands well separated from rest of the spectrum by the gaps (around $-3$~eV and $1$~eV, see Fig.~\ref{fig.band}).
Similar situation was previously reported for FeSe~\cite{ptok.kapcia.17}, where bands related to the Fe {\it d}-orbitals were well separated. 
Contrary to this, in the case of KFe$_{2}$As$_{2}$, the bands were not separated due to the strong mixing with As {\it p}-orbitals~\cite{ptok.kapcia.17}.
In the case of RbNi$_{2}$Se$_{2}$, the Se {\it p}-orbitals are mostly localized at energies from $-6$~eV to $-3$~eV, and show only a small contribution close to the Fermi energy. 
For bands above $2$~eV, the strong mixing of orbitals is observed, without dominant role of any of them.

\begin{figure}[!b]
\centering
\includegraphics[width=0.5\linewidth]{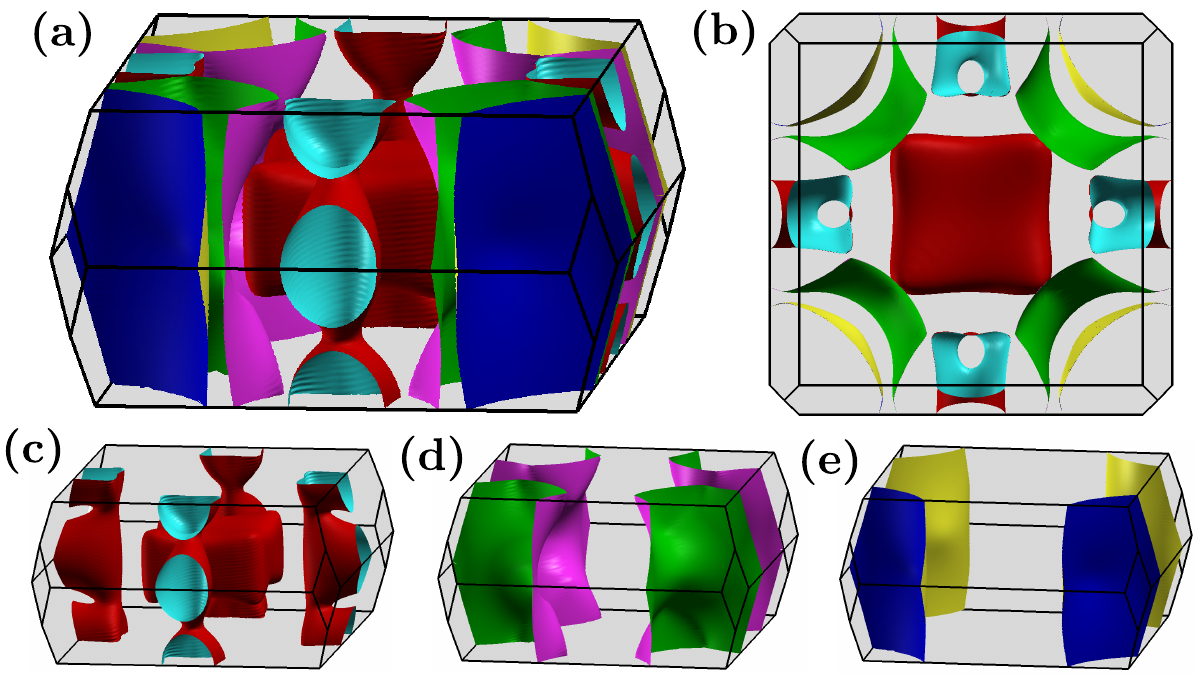}
\caption{The general view (a) and view from the top (b) of the Fermi surface of RbNi$_{2}$Se$_{2}$.
Panels (c)--(e) present the separated Fermi pockets.}
\label{fig.fs}
\end{figure}

In relation to the 122 iron-based superconductors~\cite{ptok.sternik.19}, the Fermi level is shifted to the higher energies.
As a consequence, the strong modification of the Fermi surface with respect to the 122 iron-based systems is observed (Fig.~\ref{fig.fs}).
There are three Fermi pockets with hole-like character.
One pocket exhibits exactly three dimensional character [Fig.~\ref{fig.fs}(c)], while two pockets along the X--P path exhibit quasi-2D character [Fig.~\ref{fig.fs}(d) and (e)]. It is worth mentioning that in the case of KFe$_{2}$As$_{2}$, the quasi-2D pockets are centered along the $\Gamma$--Z path, while along the X--P path there are four small pockets~\cite{ptok.sternik.19}.

The multiband character of this compound can affect its superconducting properties.
Initial investigations of the superconducting states predicted two-gap BCS-like superconductivity~\cite{liu.hu.22}.
However, RbNi$_{2}$Se$_{2}$ exhibits a weak ferromagnetic sign from the hysteresis loop, which can be ascribed to a small amount of Ni impurity~\cite{liu.hu.22}.
Nevertheless, similarly to the 122 iron-based superconductors, the magnetic instability can be expected in this system.
As a consequence, the superconducting gap can exhibit non-uniform properties, like $s_{\pm}$-wave symmetry in iron-based compounds~\cite{hirschfeld.korshunov.11}.
Unfortunately, the specific heat study typically does not give unequivocal settlement about the gap character~\cite{ptok.kapcia.20f}.

\begin{figure}[!t]
\centering
\includegraphics[width=0.6\linewidth]{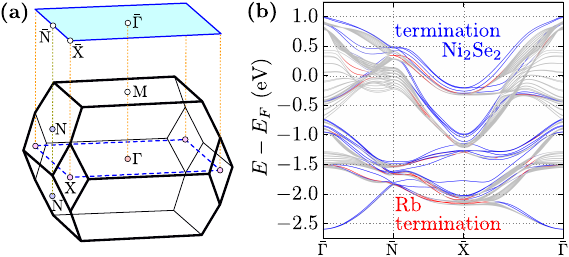}
\caption{(a) Projection of the bulk Brillouin zone on the surface Brillouin zone for the I4/mmm symmetry.
(b) The slab band structure for the (001) surface of RbNi$_{2}$Se$_{2}$ 
(the bulk states are marked by grey lines, while surface states for Ni$_{2}$Se$_{2}$ and Rb termination, are marked by blue and red lines, respectively).
}
\label{fig.slab}
\end{figure}

\begin{figure}[!b]
\centering
\includegraphics[width=0.7\linewidth]{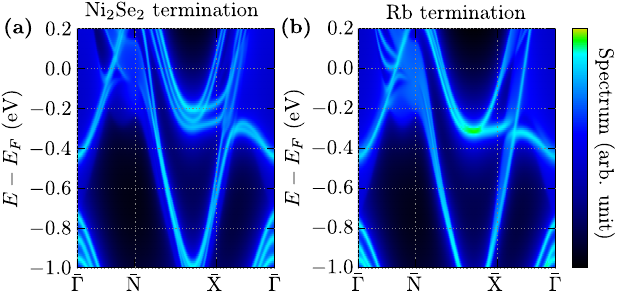}
\caption{Spectral function for the (001) surface of RbNi$_{2}$Se$_{2}$ (with different surface terminations, as labeled).}
\label{fig.spec}
\end{figure}

\subsection{Surface states}
\label{sec.surf}

Many of the iron-based superconductors exhibit topological properties~\cite{ma.wang.22}.
Here also, investigation of the electronic band structure indicated the important role of the spin--orbit coupling (described in earlier paragraphs).

\begin{figure}[!t]
\centering
\includegraphics[width=0.6\linewidth]{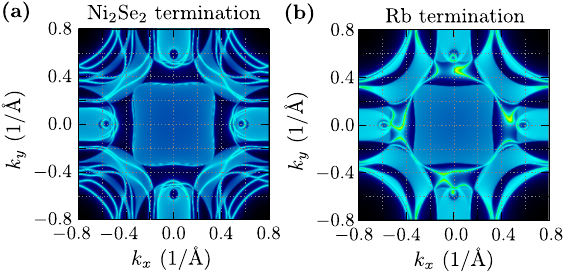}
\caption{The crossection of spectral function for the (001) surface of RbNi$_{2}$Se$_{2}$ at the Fermi level (with different surface terminations, as labeled).}
\label{fig.econst}
\end{figure}

Using the slab type calculations, we studied the termination dependence of the surface states (Fig.~\ref{fig.slab}).
We found that the surface states are realized in the system, independently on the terminated surface (Rb or Ni$_{2}$Se$_{2}$).
For both the terminations, surface states crossing the Fermi level should be visible in the angle-resolved photoemission spectroscopy (ARPES) measurements.
To investigate it more closely, we calculated the spectral function for the (001) surface (Fig.~\ref{fig.spec}).
The Rb and Ni$_{2}$Se$_{2}$ terminations should be distinguishable in high quality ARPES measurements -- the strong differences are visible along the $\bar{\text{N}}$--$\bar{\text{X}}$ path, i.e. along the edge of 2D Brillouin zone [cf. Fig.~\ref{fig.spec}(a) and~\ref{fig.spec}(b)].
This is reflected also in the isoenergetic crossection of the spectral function (Fig.~\ref{fig.econst}).
As we can see, the 2D energy contour at the Fermi level exhibits strong differences, while the surface states are well visible for the Ni$_{2}$Se$_{2}$ termination.
Comparing our theoretical results with the experimental ones (discussed in Ref.~\cite{liu.hu.22}), we can conclude that the experimentally investigated sample had the Rb termination.
The Ni$_{2}$Se$_{2}$ terminated sample should realize several well visible bands (e.g. along the $\bar{\Gamma}$--$\bar{\text{N}}$ direction).
Contrary to this, for the surface with the Rb termination, the energy contour are mostly associated with the bulk-like states (blurred spectrum).
Nevertheless, few strong surface states are also visible around the $\bar{\Gamma}$ point, in the middle of the 2D Brillouin zone.

\begin{figure}[!b]
\centering
\includegraphics[width=0.6\linewidth]{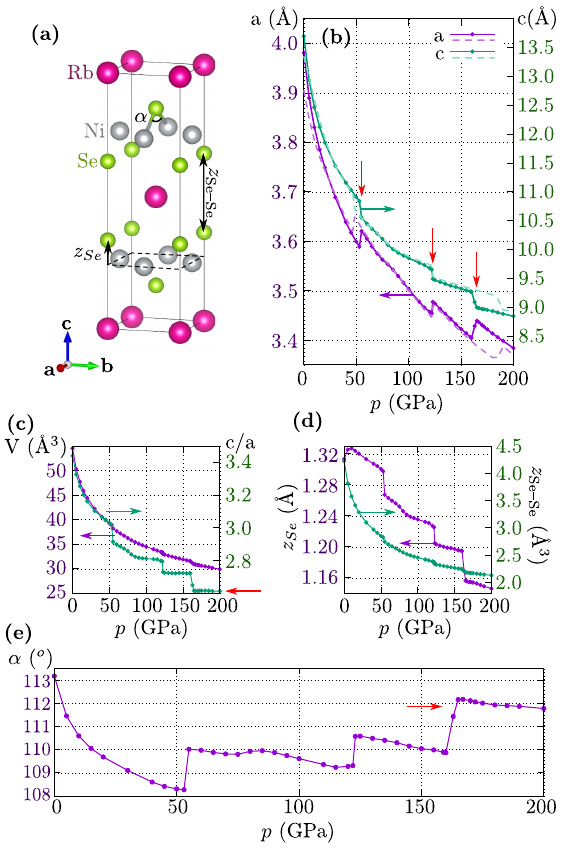}
\caption{(a) The conventional cell of the I4/mmm structure of RbNi$_{2}$Se$_{2}$.
Pressure dependence of the following parameters: (b) lattice constants $a$ and $c$, (c) volume of the conventional cell $V$ and $c/a$ ratio, (d) distance $z_\text{Se}$ of Se atom from the Ni plane and distance $z_\text{Se--Se}$ between Se atoms along the $c$ direction [shown in (a)], and (e) angle between Ni--Se bonds [shown in (a)].
On panel (b), we compare also lattice constant $a$ and $c$  obtained from calculation with ``standard'' PAW potentials (solid line), and PAW potentials containing semicore electrons as valence one (dashed line).
}
\label{fig.press}
\end{figure}

\subsection{Role of hydrostatic pressure}
\label{sec.press}

As we mentioned in the introduction, the external hydrostatic pressure can lead to the isostructural phase transition from the tetragonal to collapsed-tetragonal phase (T-cT transition), which was reported in many 122-type compounds~\cite{kreyssig.green.08,uhoya.stemshorn.10,mittal.mishra.11,uhoya.tsoi.10,tafti.juneaufecteau.13,taufour.foroozani.14,tafti.clancy.14,tafti.ouellet.15,wang.matsubayashi.16,ptok.sternik.19,ptok.kapcia.20}.
In case of RhNi$_{2}$Se$_{2}$ similar properties should be expected.
Surprisingly, our study uncover a much complicated situation (Fig.~\ref{fig.press}), with multiple isostructural phase transitions.

Now we compare the modification of the structural parameters under the external hydrostatic pressure, obtained for RbNi$_{2}$Se$_{2}$ (Fig.~\ref{fig.press}) and KFe$_{2}$As$_{2}$~\cite{ptok.sternik.19}:
\begin{itemize}
\item Under pressure, we observe simultaneous discontinuous decreasing of $a$ and $c$ [Fig.~\ref{fig.press}(b)] -- several isostructural phase transitions are visible (marked by red arrows).
In these transitions, the lattice constant $a$ increases by $\delta a \simeq 0.03$~\AA, while $c$ decreases by $\delta c \simeq 0.3$~\AA.
This behavior can be compared to the case of KFe$_{2}$As$_{2}$, where the lattice constant modifications were much stronger: $\delta a \simeq 0.1$~\AA\ and $\delta c \simeq 1$~\AA~\cite{ptok.sternik.19}.
Additionally, in the KFe$_{2}$As$_{2}$ compound, $a$ increases slightly just before the final collapse of the cell at critical pressure.
In case of RbNi$_{2}$Se$_{2}$, the lattice constant $a$ increases only at the transition pressures.
\item As a natural consequence of the external hydrostatic pressure, the volume $V$ of the unit cell decreases, while the $c/a$ ratio shows discontinuous decreasing character [Fig.~\ref{fig.press}(c)].
During the isostructural transitions, the volume does not exhibit dramatical changes.
Additionally, with pressure, the $c/a$ ratio goes to the saturated value $\sim 2.6$ (marked by red arrow).
\item Modification of the lattice constants indicates a change in the distance between atoms [Fig.~\ref{fig.press}(d)].
The interlayer distance between two Se atoms [labeled as $z_\text{Se--Se}$ in Fig.~\ref{fig.press}(a)] decreases slowly.
Simultaneously the distance of Se atoms from the Ni square net plane [labeled as $z_\text{Se}$ in Fig.~\ref{fig.press}(a)] drops dramatically a few times.
The distance $z_\text{Se}$ shows a similar behavior like the distance between As atoms and Fe square net plane in KFe$_{2}$As$_{2}$~\cite{ptok.sternik.19} -- the $z_\text{Se}$ drops at the pressures related to the isostructural transitions.
\item The interplanar Se-Ni-Se bond angle [marked in Fig~\ref{fig.press}(a) as $\alpha$] shows a nonmonotonous dependence on pressure and changes its value a several times [Fig.~\ref{fig.press}(e)].
Between the transitions, $\alpha$ for $p > 50$~GPa has approximately constant value, while for extremely large pressure, its value saturates to $\alpha \simeq 112^{o}$. 
For 122 iron-based superconductors, $\alpha$ increases with pressure to the saturated value $116^{o}$.
\end{itemize}
Here we would like to mention that the initial lattice constants (in the absence of pressure), as well as the $c/a$ ration, for RbNi$_{2}$Se$_{2}$ and KFe$_{2}$As$_{2}$ were comparable.
Despite of it, the properties of these systems under external hydrostatic pressure do not posses many similarities.
However, our prediction of the multiple phase transitions in RbNi$_2$Se$_2$ could be verified experimentally using the diffraction method.

In context of the superconducting properties of RbNi$_{2}$Se${2}$ under pressure, the $z_\text{Se}$ parameter can play an important role.
In the iron-based superconductors, the distance between the anion and the Fe plane can be related to the observed $T_{c}$~\cite{mizuguchi.hara.10,kang.birol.17}.
A decrease in the value of $z_\text{Se}$ can suggest a destructive role of pressure on the superconducting phase.

\begin{figure}[!tp]
\centering
\includegraphics[width=0.975\columnwidth]{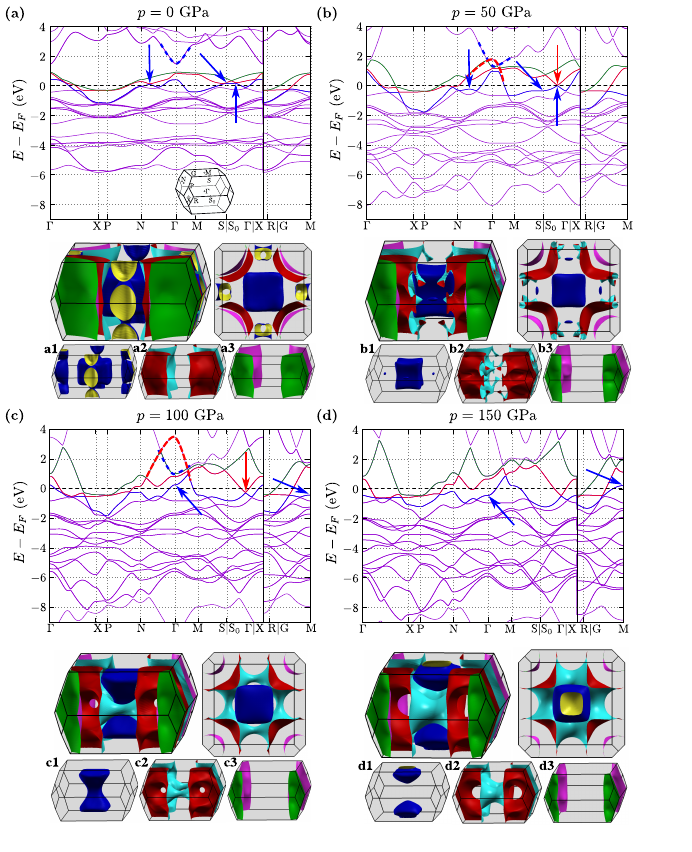}
\caption{Electronic band structure (upper panels) and corresponding Fermi surface (lower panels) of RbNi$_{2}$Se$_{2}$ under external hydrostatic pressure (as labeled).
For guide for the eye the bands crossing the Fermi level, are marked by red, green, and blue color (related to the color of the separated Fermi pocket on panels a1, a2, a3, etc.).
}
\label{fig.el_press}
\end{figure}

Recent studies of the system under pressure indicate the important role of the electrons, which typically are treated as the core ones~\cite{zhang.wang.21}.
Under high pressure, the electronic structure of a system can change significantly compared to that at ambient pressure.
One effect that has been recognized is that the transfer of conduction electrons between the {\it s}-{\it p} states and {\it d} states depends on the degree of compression and varies among crystal structures~\cite{pettifor.77,moriarty.92,grad.blaha.00}.
This property can be important in context of the system stability, due to the impact of the conduction electrons.
The ``standard'' PAW potential can give erroneous thermodynamic properties under high pressure.
Therefore, the PAW potential containing the semicore electrons are also available, and can improve the obtained results.
Indeed, for alkali metal Rb, two versions of VASP PAW potentials can be used (containing 7 and 9 valence electrons -- both containing the semicore states, {\it s} and {\it p}, respectively), and similarly for the transition metal Ni (one PAW potential without semicore electrons, containing 10 valence electrons, and second PAW potential with {\it p} semicore electrons, containing 16 valence electrons).
To check quality of obtained results, we compare the lattice constant of the RbNi$_{2}$Se$_{2}$ under pressure (Fig.~\ref{sec.press}(b)) -- results obtained with ``standard'' PAW potentials (containing 82 valence electrons within the conventional cell) are presented by dots and solid line, while results for PAW potentials with semicore electrons (corresponding 106 valence electrons within the conventional cell) by dashed line.
As we can see, both type PAW potentials reproduce the same behavior of $a$ and $c$ under pressure (also few ``jumps'' of lattice constant is visible in Fig.~\ref{fig.press}(a)). 
However, under extremely high pressure (above $150$~GPa) the results are significantly different -- jump of $a$ and $c$ is shifted to higher pressure.
Nevertheless, we should have in mind that even if the presented results can correctly reproduce experimentally observed $a$ and $c$ behavior (like earlier study of KFe$_{2}$As$_{2}$~\cite{ptok.sternik.19}), it should be used with the uttermost care in the presence of extremely high-pressure calculations.

\paragraph*{Electronic properties and Lifshitz transition under pressure.}
The external hydrostatic pressure also have an impact on the electronic band structure.
The modified distances between the atoms lead to a stronger overlap between the atomic orbitals.
As a consequence, the electronic band structure as well as the Fermi surface can be strongly modified.

Firstly, the bandwidth increases -- this is well visible when we compare the electronic band structure at different pressure (Fig.~\ref{fig.el_press}). 
For example at the $\Gamma$ point, the electron-like parabolic band (marked by blue dashed lines) is shifted to lower energies, while at the same time, the hole-like band (marked by red dashed lines) is shifted to higher energies. 
Similar behavior was reported previously in context of the transition from the tetragonal to collapsed-tetragonal phase in KFe$_{2}$As$_{2}$~\cite{ptok.sternik.19}, where the most notable modifications were observed within the $p_{z}$ orbitals of anions.

Secondly, in some situations, specific external conditions can lead to a change of the Fermi surface topology.
Such modification of the Fermi surface, known as the {\it Lifshitz transition}~\cite{lifshitz.60}, can be induced by doping~\cite{norman.lin.10,lebouf.boironleyraund.11,sato.nakayama.09,nakayama.sato.11,maleab.shimojima.12,xu.richard.13,hodovanets.liu.14,liu.lograsson.14,cho.konczykowski.16,khan.johnson.14,benhabib.sacuto.15,liu.kondo.10}, external pressure~\cite{harrison.sebastian.07,purcell.graf.09}, or magnetic field~\cite{daou.bergemann.06,schlottmann.11,ptok.cichy.17,ptok.kapcia.17m,ptok.18}.
In the case of 122 iron based superconductors (e.g., Ba$_{1-x}$K$_{x}$Fe$_{2}$As$_{2}$~\cite{xu.richard.13,maleab.shimojima.12,nakayama.sato.11,sato.nakayama.09,cho.konczykowski.16,liu.lograsson.14,hodovanets.liu.14,khan.johnson.14} or Ba(Fe$_{1-x}$Co$_{x}$)$_{2}$As$_{2}$~\cite{liu.kondo.10}), strong dependence of the Fermi surface on doping was observed experimentally.
Also, the pressure dependent Lifshitz transition in the iron-based systems was investigated theoretically~\cite{ptok.sternik.19,ptok.kapcia.20,tresca.profeta.17,sen.guo.20,ghosh.sen.21,ghosh.ghosh.22}.
In our case, the pressure also leads to a Lifshitz transition (cf. the Fermi surfaces presented in Fig.~\ref{fig.el_press}).

As we mentioned in Sec.~\ref{fig.band}, at the ambient pressure, the Fermi surface is composed of three separated pockets 
(Fig.~\ref{fig.el_press}(a1)-(a3)).
Number of the packets does not depend on pressure (in this part, we will call it first, second, and thirt pocket, what will be corresponding to pockets in Figs.~\ref{fig.el_press}(a1), (a2), and (a3), respectively). 
The external pressure affects the first Fermi pocket the most (Fig.~\ref{fig.el_press}(a1)).
In the ambient pressure, this pocket exhibits the most three-dimensional character (with the strong $k_{z}$-dependence dispersion).
In the first stage, increasing pressure leads to the disappearance of part of the pocket along the S$_{0}$--N--S path.
The remnant of this part is a small ellipsoidal-like shape along S$_{0}$--$\Gamma$ at $50$~GPa (blue arrow in Fig.~\ref{fig.band}(b)).
Nevertheless, this pocket is mainly realized by a cubic-like shape centered at the $\Gamma$ point.
This modification of the first pocket is related to the Fermi level by the bands around the N and S$_{0}$ points (cf.~blue arrows in Fig.~\ref{fig.el_press}(a) and (b)).
Further increase of pressure leads to a modification of the cubic-like pocket (cf.~Fig.~\ref{fig.el_press}(b1), (c1), and (d1)).
For $p > 50$~GPa, the band at the $\Gamma$ point is shifted to the lower energy, while the one at the M point is shifted to the higher energy (cf.~blue arrows in Fig.~\ref{fig.el_press}(c) and (d)).
During this transition, the first pocket changes shape from cubic-like (Fig.~\ref{fig.el_press}(b1)), through dumbbell-like along $k_{z}$ and centered at the $\Gamma$ point (Fig.~\ref{fig.el_press}(c1)), to bowtie-like along $k_{z}$ and centered at the M point (Fig.~\ref{fig.el_press}(d1)).
The second Fermi pocket, with an initially cylindrical shape (~Fig.~\ref{fig.el_press}(a2)), acquires new features -- it changes character from quasi-two-dimensional to directly three-dimensional with the strong $k_{z}$-dependence dispersion.
This is related to crossing the Fermi level by the bands along the $\Gamma$--S$_{0}$ path -- the bottom of band crosses the Fermi level for pressure between $50$~GPa and $100$~GPa (cf.~red arrows on Fig.~\ref{fig.el_press}(b) and (c)).
Further increase of pressure does not introduce any bigger modification of this pocket shape.
The third Fermi pocket does not change its shape under pressure (cf.~Fig.~\ref{fig.el_press}(a3), (b3), (c3), and (d3)). 
The cylindrical shape of this pocket is related to the quasi-two-dimensional character of the band, which constitutes it (parabolic band along the X--P path with the weak $k_{z}$-dependence dispersion).

\section{Summary}
\label{sec.summary}

In conclusions, we investigated the  electronic properties and surface states of RbNi$_{2}$Se$_{2}$~\cite{li.li.12}.
The obtained Fermi surface suggests quasi-two-dimensional character of the electronic states in the two bands crossing the Fermi level, while the third band has fully three dimensional character.
We presented the electronic band structure in the absence and presence of the spin--orbit coupling.
There are a few places in the band structure, where the spin--orbit coupling plays an important role. 
As a consequence of the topological effects, the surface states can emerge independent of the surface terminations.
However, the spectral function depends on the type of termination (Rb or Ni$_{2}$Se$_{2}$ termination).
Indeed, we show that the ARPES results presented in Ref.~\cite{li.li.12} suggest that the sample had the Rb termination.
Moreover, in case of the Ni$_{2}$Se$_{2}$ termination, the electronic band structure posses much more complex structure, due to the realization of many surface states crossing the Fermi level.

Additionally, we discussed the multiple isostructural phase transitions under external hydrostatic pressure. 
Contrary to the 122 type iron-based superconductors, RbNi$_{2}$Se$_{2}$ does not exhibit transition from the tetragonal to collapsed-tetragonal phase. 
In case of RbNi$_{2}$Se$_{2}$, internal chemical pressure balances the external chemical pressure and the collapse of tetragonal cell does not occur.
Nevertheless, modification of the distance between atoms leads to a change in the electronic band structure as a consequence of an increased overlap between the atomic orbitals.
This band structure modification is reflected in the Fermi surface features, which exhibit the Lifshitz transition -- the Fermi pockets show strong pressure dependence.

\section*{Acknowledgments}

Some figures in this work were rendered using {\sc Vesta}~\cite{momma.izumi.11} and {\sc XCrySDen}~\cite{kokalj.99} softwares.
S.B. is grateful to IT4Innovations (V\v{S}B-TU Ostrava) for hospitality during a part of the work on this project. 
This work was supported by National Science Centre (NCN, Poland) under Projects No.
2017/25/B/ST3/02586 (S.B.) 
and
2021/43/B/ST3/02166 (P.P. and A.P.). 
A.P. appreciates funding in the frame of scholarships of the 
Minister of Science and Higher Education (Poland) for outstanding young 
scientists (2019 edition, No. 818/STYP/14/2019).

\printcredits

\section*{Declaration of Competing Interest}

The authors declare that they have no known competing financial interests or personal relationships that could have appeared to influence the work reported in this paper.

\section*{Data Availability Statement}

The data presented in this study are available on request from the authors.


\bibliographystyle{spphys}
\bibliography{biblio.bib}

\begin{thebibliography}{10}
\providecommand{\url}[1]{{#1}}
\providecommand{\urlprefix}{URL }
\expandafter\ifx\csname urlstyle\endcsname\relax
  \providecommand{\doi}[1]{DOI \discretionary{}{}{}#1}\else
  \providecommand{\doi}{DOI \discretionary{}{}{}\begingroup
  \urlstyle{rm}\Url}\fi

\bibitem{kamihara.watanabe.08}
Y.~Kamihara, T.~Watanabe, M.~Hirano, H.~Hosono, J. Am. Chem. Soc. \textbf{130},
  3296 (2008).
\newblock \doi{10.1021/ja800073m}.
\newblock \urlprefix\url{https://doi.org/10.1021/ja800073m}

\bibitem{ren.zhao.09}
Z.A. Ren, Z.X. Zhao, Adv. Mater. \textbf{21}, 4584 (2009).
\newblock \doi{10.1002/adma.200901049}.
\newblock \urlprefix\url{https://doi.org/10.1002/adma.200901049}

\bibitem{stewart.11}
G.R. Stewart, Rev. Mod. Phys. \textbf{83}, 1589 (2011).
\newblock \doi{10.1103/RevModPhys.83.1589}.
\newblock \urlprefix\url{https://doi.org/10.1103/RevModPhys.83.1589}

\bibitem{kordyuk.12}
A.A. Kordyuk, Low Temp. Phys. \textbf{38}, 888 (2012).
\newblock \doi{10.1063/1.4752092}.
\newblock \urlprefix\url{https://doi.org/10.1063/1.4752092}

\bibitem{hosono.kuroki.15}
H.~Hosono, K.~Kuroki, Phys. C \textbf{514}, 399 (2015).
\newblock \doi{https://doi.org/10.1016/j.physc.2015.02.020}.
\newblock \urlprefix\url{https://doi.org/10.1016/j.physc.2015.02.020}

\bibitem{fernandes.chubukov.16}
R.M. Fernandes, A.V. Chubukov, Reports on Progress in Physics \textbf{80},
  014503 (2016).
\newblock \doi{10.1088/1361-6633/80/1/014503}.
\newblock \urlprefix\url{https://doi.org/10.1088/1361-6633/80/1/014503}

\bibitem{luo.wang.08}
H.~Luo, Z.~Wang, H.~Yang, P.~Cheng, X.~Zhu, H.H. Wen, Supercond. Sci. Technol.
  \textbf{21}, 125014 (2008).
\newblock \doi{10.1088/0953-2048/21/12/125014}.
\newblock \urlprefix\url{https://doi.org/10.1088/0953-2048/21/12/125014}

\bibitem{neupane.richard.11}
M.~Neupane, P.~Richard, Y.M. Xu, K.~Nakayama, T.~Sato, T.~Takahashi, A.V.
  Federov, G.~Xu, X.~Dai, Z.~Fang, Z.~Wang, G.F. Chen, N.L. Wang, H.H. Wen,
  H.~Ding, Phys. Rev. B \textbf{83}, 094522 (2011).
\newblock \doi{10.1103/PhysRevB.83.094522}.
\newblock \urlprefix\url{https://doi.org/10.1103/PhysRevB.83.094522}

\bibitem{rotter.tegel.08}
M.~Rotter, M.~Tegel, D.~Johrendt, I.~Schellenberg, W.~Hermes, R.~P\"ottgen,
  Phys. Rev. B \textbf{78}, 020503 (2008).
\newblock \doi{10.1103/PhysRevB.78.020503}.
\newblock \urlprefix\url{https://doi.org/10.1103/PhysRevB.78.020503}

\bibitem{rotter.tegel.08b}
M.~Rotter, M.~Tegel, D.~Johrendt, Phys. Rev. Lett. \textbf{101}, 107006 (2008).
\newblock \doi{10.1103/PhysRevLett.101.107006}.
\newblock \urlprefix\url{https://doi.org/10.1103/PhysRevLett.101.107006}

\bibitem{tegel.marianne.08}
M.~Tegel, M.~Rotter, V.~Wei{\ss}, F.M. Schappacher, R.~P\"{o}ttgen,
  D.~Johrendt, J. Phys.: Condens. Matter \textbf{20}, 452201 (2008).
\newblock \doi{10.1088/0953-8984/20/45/452201}.
\newblock \urlprefix\url{https://doi.org/10.1088/0953-8984/20/45/452201}

\bibitem{shirage.miyazawa.08}
P.M. Shirage, K.~Miyazawa, H.~Kito, H.~Eisaki, A.~Iyo, Appl. Phys. Express
  \textbf{1}, 081702 (2008).
\newblock \doi{10.1143/apex.1.081702}.
\newblock \urlprefix\url{https://doi.org/10.1143/apex.1.081702}

\bibitem{anupam.paulose.09}
Anupam, P.L. Paulose, H.S. Jeevan, C.~Geibel, Z.~Hossain, J. Phys.: Condens.
  Matter \textbf{21}, 265701 (2009).
\newblock \doi{10.1088/0953-8984/21/26/265701}.
\newblock \urlprefix\url{https://doi.org/10.1088/0953-8984/21/26/265701}

\bibitem{alireza.alireza.ko.08}
P.L. Alireza, Y.T.C. Ko, J.~Gillett, C.M. Petrone, J.M. Cole, G.G. Lonzarich,
  S.E. Sebastian, J. Phys.: Condens. Matter \textbf{21}, 012208 (2008).
\newblock \doi{10.1088/0953-8984/21/1/012208}.
\newblock \urlprefix\url{https://doi.org/10.1088/0953-8984/21/1/012208}

\bibitem{huang.qiu.08}
Q.~Huang, Y.~Qiu, W.~Bao, M.A. Green, J.W. Lynn, Y.C. Gasparovic, T.~Wu, G.~Wu,
  X.H. Chen, Phys. Rev. Lett. \textbf{101}, 257003 (2008).
\newblock \doi{10.1103/PhysRevLett.101.257003}.
\newblock \urlprefix\url{https://doi.org/10.1103/PhysRevLett.101.257003}

\bibitem{leithejasper.schnelle.08}
A.~Leithe-Jasper, W.~Schnelle, C.~Geibel, H.~Rosner, Phys. Rev. Lett.
  \textbf{101}, 207004 (2008).
\newblock \doi{10.1103/PhysRevLett.101.207004}.
\newblock \urlprefix\url{https://doi.org/10.1103/PhysRevLett.101.207004}

\bibitem{harrison.mcdonald.09}
N.~Harrison, R.D. McDonald, C.H. Mielke, E.D. Bauer, F.~Ronning, J.D. Thompson,
  J. Phys.: Condens. Matter \textbf{21}, 322202 (2009).
\newblock \doi{10.1088/0953-8984/21/32/322202}.
\newblock \urlprefix\url{https://doi.org/10.1088/0953-8984/21/32/322202}

\bibitem{baek.lee.09}
S.H. Baek, H.~Lee, S.E. Brown, N.J. Curro, E.D. Bauer, F.~Ronning, T.~Park,
  J.D. Thompson, Phys. Rev. Lett. \textbf{102}, 227601 (2009).
\newblock \doi{10.1103/PhysRevLett.102.227601}.
\newblock \urlprefix\url{https://doi.org/10.1103/PhysRevLett.102.227601}

\bibitem{shirage.kihou.09}
P.M. Shirage, K.~Kihou, K.~Miyazawa, C.H. Lee, H.~Kito, H.~Eisaki,
  T.~Yanagisawa, Y.~Tanaka, A.~Iyo, Phys. Rev. Lett. \textbf{103}, 257003
  (2009).
\newblock \doi{10.1103/PhysRevLett.103.257003}.
\newblock \urlprefix\url{https://doi.org/10.1103/PhysRevLett.103.257003}

\bibitem{kawashima.ishida.18}
K.~Kawashima, S.~Ishida, H.~Fujihisa, Y.~Gotoh, Y.~Yoshida, H.~Eisaki,
  H.~Ogino, A.~Iyo, Sci. Rep. \textbf{8}, 16827 (2018).
\newblock \doi{10.1038/s41598-018-34265-2}.
\newblock \urlprefix\url{https://doi.org/10.1038/s41598-018-34265-2}

\bibitem{kreyssig.green.08}
A.~Kreyssig, M.A. Green, Y.~Lee, G.D. Samolyuk, P.~Zajdel, J.W. Lynn, S.L.
  Bud'ko, M.S. Torikachvili, N.~Ni, S.~Nandi, J.B. Le\~ao, S.J. Poulton, D.N.
  Argyriou, B.N. Harmon, R.J. McQueeney, P.C. Canfield, A.I. Goldman, Phys.
  Rev. B \textbf{78}, 184517 (2008).
\newblock \doi{10.1103/PhysRevB.78.184517}.
\newblock \urlprefix\url{https://doi.org/10.1103/PhysRevB.78.184517}

\bibitem{uhoya.stemshorn.10}
W.~Uhoya, A.~Stemshorn, G.~Tsoi, Y.K. Vohra, A.S. Sefat, B.C. Sales, K.M. Hope,
  S.T. Weir, Phys. Rev. B \textbf{82}, 144118 (2010).
\newblock \doi{10.1103/PhysRevB.82.144118}.
\newblock \urlprefix\url{https://doi.org/10.1103/PhysRevB.82.144118}

\bibitem{mittal.mishra.11}
R.~Mittal, S.K. Mishra, S.L. Chaplot, S.V. Ovsyannikov, E.~Greenberg, D.M.
  Trots, L.~Dubrovinsky, Y.~Su, T.~Brueckel, S.~Matsuishi, H.~Hosono,
  G.~Garbarino, Phys. Rev. B \textbf{83}, 054503 (2011).
\newblock \doi{10.1103/PhysRevB.83.054503}.
\newblock \urlprefix\url{https://doi.org/10.1103/PhysRevB.83.054503}

\bibitem{uhoya.tsoi.10}
W.~Uhoya, G.~Tsoi, Y.K. Vohra, M.A. McGuire, A.S. Sefat, B.C. Sales,
  D.~Mandrus, S.T. Weir, J. Phys.: Condens. Matter \textbf{22}, 292202 (2010).
\newblock \doi{10.1088/0953-8984/22/29/292202}.
\newblock \urlprefix\url{https://doi.org/10.1088/0953-8984/22/29/292202}

\bibitem{tafti.juneaufecteau.13}
F.F. Tafti, A.~Juneau-Fecteau, M.{\`E}. Delage, S.~Ren{\'e}~de Cotret, J.P.
  Reid, A.F. Wang, X.G. Luo, X.H. Chen, N.~Doiron-Leyraud, L.~Taillefer, Nat.
  Phys. \textbf{9}, 349 (2013).
\newblock \doi{10.1038/nphys2617}.
\newblock \urlprefix\url{https://doi.org/10.1038/nphys2617}

\bibitem{taufour.foroozani.14}
V.~Taufour, N.~Foroozani, M.A. Tanatar, J.~Lim, U.~Kaluarachchi, S.K. Kim,
  Y.~Liu, T.A. Lograsso, V.G. Kogan, R.~Prozorov, S.L. Bud'ko, J.S. Schilling,
  P.C. Canfield, Phys. Rev. B \textbf{89}, 220509 (2014).
\newblock \doi{10.1103/PhysRevB.89.220509}.
\newblock \urlprefix\url{https://doi.org/10.1103/PhysRevB.89.220509}

\bibitem{tafti.clancy.14}
F.F. Tafti, J.P. Clancy, M.~Lapointe-Major, C.~Collignon, S.~Faucher, J.A.
  Sears, A.~Juneau-Fecteau, N.~Doiron-Leyraud, A.F. Wang, X.G. Luo, X.H. Chen,
  S.~Desgreniers, Y.J. Kim, L.~Taillefer, Phys. Rev. B \textbf{89}, 134502
  (2014).
\newblock \doi{10.1103/PhysRevB.89.134502}.
\newblock \urlprefix\url{https://doi.org/10.1103/PhysRevB.89.134502}

\bibitem{tafti.ouellet.15}
F.F. Tafti, A.~Ouellet, A.~Juneau-Fecteau, S.~Faucher, M.~Lapointe-Major,
  N.~Doiron-Leyraud, A.F. Wang, X.G. Luo, X.H. Chen, L.~Taillefer, Phys. Rev. B
  \textbf{91}, 054511 (2015).
\newblock \doi{10.1103/PhysRevB.91.054511}.
\newblock \urlprefix\url{https://doi.org/10.1103/PhysRevB.91.054511}

\bibitem{wang.matsubayashi.16}
B.~Wang, K.~Matsubayashi, J.~Cheng, T.~Terashima, K.~Kihou, S.~Ishida, C.H.
  Lee, A.~Iyo, H.~Eisaki, Y.~Uwatoko, Phys. Rev. B \textbf{94}, 020502 (2016).
\newblock \doi{10.1103/PhysRevB.94.020502}.
\newblock \urlprefix\url{https://doi.org/10.1103/PhysRevB.94.020502}

\bibitem{ptok.sternik.19}
A.~Ptok, M.~Sternik, K.J. Kapcia, P.~Piekarz, Phys. Rev. B \textbf{99}, 134103
  (2019).
\newblock \doi{10.1103/PhysRevB.99.134103}.
\newblock \urlprefix\url{https://doi.org/10.1103/PhysRevB.99.134103}

\bibitem{ptok.kapcia.20}
A.~Ptok, K.J. Kapcia, M.~Sternik, P.~Piekarz, J. Supercond. Nov. Magn.
  \textbf{33}, 2347 (2020).
\newblock \doi{10.1007/s10948-020-05454-w}.
\newblock \urlprefix\url{https://doi.org/10.1007/s10948-020-05454-w}

\bibitem{tresca.profeta.17}
C.~Tresca, G.~Profeta, Phys. Rev. B \textbf{95}, 165129 (2017).
\newblock \doi{10.1103/PhysRevB.95.165129}.
\newblock \urlprefix\url{https://doi.org/10.1103/PhysRevB.95.165129}

\bibitem{liu.hu.22}
H.~Liu, X.~Hu, H.~Guo, X.K. Teng, H.~Bu, Z.~Luo, L.~Li, Z.~Liu, M.~Huo,
  F.~Liang, H.~Sun, B.~Shen, P.~Dai, R.J. Birgeneau, D.X. Yao, M.~Yi, M.~Wang,
  Phys. Rev. B \textbf{106}, 094511 (2022).
\newblock \doi{10.1103/PhysRevB.106.094511}.
\newblock \urlprefix\url{https://doi.org/10.1103/PhysRevB.106.094511}

\bibitem{gurevich.11}
A.~Gurevich, Rep. Prog. Phys. \textbf{74}, 124501 (2011).
\newblock \doi{10.1088/0034-4885/74/12/124501}.
\newblock \urlprefix\url{https://doi.org/10.1088/0034-4885/74/12/124501}

\bibitem{neilson.llobet.12}
J.R. Neilson, A.~Llobet, A.V. Stier, L.~Wu, J.~Wen, J.~Tao, Y.~Zhu, Z.B.
  Tesanovic, N.P. Armitage, T.M. McQueen, Phys. Rev. B \textbf{86}, 054512
  (2012).
\newblock \doi{10.1103/PhysRevB.86.054512}.
\newblock \urlprefix\url{https://doi.org/10.1103/PhysRevB.86.054512}

\bibitem{chen.yang.16}
H.~Chen, J.~Yang, C.~Cao, L.~Li, Q.~Su, B.~Chen, H.~Wang, Q.~Mao, B.~Xu, J.~Du,
  M.~Fang, Supercond. Sci. Technol. \textbf{29}, 045008 (2016).
\newblock \doi{10.1088/0953-2048/29/4/045008}.
\newblock \urlprefix\url{https://doi.org/10.1088/0953-2048/29/4/045008}

\bibitem{wang.dong.13}
H.~Wang, C.~Dong, Q.~Mao, R.~Khan, X.~Zhou, C.~Li, B.~Chen, J.~Yang, Q.~Su,
  M.~Fang, Phys. Rev. Lett. \textbf{111}, 207001 (2013).
\newblock \doi{10.1103/PhysRevLett.111.207001}.
\newblock \urlprefix\url{https://doi.org/10.1103/PhysRevLett.111.207001}

\bibitem{giannozzi.baroni.09}
P.~Giannozzi, S.~Baroni, N.~Bonini, M.~Calandra, R.~Car, C.~Cavazzoni,
  D.~Ceresoli, G.L. Chiarotti, M.~Cococcioni, I.~Dabo, A.D. Corso,
  S.~de~Gironcoli, S.~Fabris, G.~Fratesi, R.~Gebauer, U.~Gerstmann,
  C.~Gougoussis, A.~Kokalj, M.~Lazzeri, L.~Martin-Samos, N.~Marzari, F.~Mauri,
  R.~Mazzarello, S.~Paolini, A.~Pasquarello, L.~Paulatto, C.~Sbraccia,
  S.~Scandolo, G.~Sclauzero, A.P. Seitsonen, A.~Smogunov, P.~Umari, R.M.
  Wentzcovitch, J. Phys.: Condens. Matter \textbf{21}, 395502 (2009).
\newblock \doi{10.1088/0953-8984/21/39/395502}.
\newblock \urlprefix\url{https://doi.org/10.1088/0953-8984/21/39/395502}

\bibitem{giannozzi.andreussi.17}
P.~Giannozzi, O.~Andreussi, T.~Brumme, O.~Bunau, M.B. Nardelli, M.~Calandra,
  R.~Car, C.~Cavazzoni, D.~Ceresoli, M.~Cococcioni, N.~Colonna, I.~Carnimeo,
  A.D. Corso, S.~de~Gironcoli, P.~Delugas, R.A. DiStasio, A.~Ferretti,
  A.~Floris, G.~Fratesi, G.~Fugallo, R.~Gebauer, U.~Gerstmann, F.~Giustino,
  T.~Gorni, J.~Jia, M.~Kawamura, H.Y. Ko, A.~Kokalj,
  E.~K\"{u}{\c{c}}\"{u}kbenli, M.~Lazzeri, M.~Marsili, N.~Marzari, F.~Mauri,
  N.L. Nguyen, H.V. Nguyen, A.O. de-la Roza, L.~Paulatto, S.~Ponc{\'{e}},
  D.~Rocca, R.~Sabatini, B.~Santra, M.~Schlipf, A.P. Seitsonen, A.~Smogunov,
  I.~Timrov, T.~Thonhauser, P.~Umari, N.~Vast, X.~Wu, S.~Baroni, J. Phys.:
  Condens. Matter \textbf{29}, 465901 (2017).
\newblock \doi{10.1088/1361-648x/aa8f79}.
\newblock \urlprefix\url{https://doi.org/10.1088/1361-648x/aa8f79}

\bibitem{giannozzi.baseggio.20}
P.~Giannozzi, O.~Baseggio, P.~Bonf\'{a}, D.~Brunato, R.~Car, I.~Carnimeo,
  C.~Cavazzoni, S.~de~Gironcoli, P.~Delugas, F.~Ferrari~Ruffino, A.~Ferretti,
  N.~Marzari, I.~Timrov, A.~Urru, S.~Baroni, J. Chem. Phys. \textbf{152},
  154105 (2020).
\newblock \doi{10.1063/5.0005082}.
\newblock \urlprefix\url{https://doi.org/10.1063/5.0005082}

\bibitem{pardew.burke.96}
J.P. Perdew, K.~Burke, M.~Ernzerhof, Phys. Rev. Lett. \textbf{77}, 3865 (1996).
\newblock \doi{10.1103/PhysRevLett.77.3865}.
\newblock \urlprefix\url{http://doi.org/10.1103/PhysRevLett.77.3865}

\bibitem{dalcolrso.14}
A.~{Dal Corso}, Comput. Mater. Sci. \textbf{95}, 337 (2014).
\newblock \doi{10.1016/j.commatsci.2014.07.043}.
\newblock \urlprefix\url{https://doi.org/10.1016/j.commatsci.2014.07.043}

\bibitem{marzari.mostofi.12}
N.~Marzari, A.A. Mostofi, J.R. Yates, I.~Souza, D.~Vanderbilt, Rev. Mod. Phys.
  \textbf{84}, 1419 (2012).
\newblock \doi{10.1103/RevModPhys.84.1419}.
\newblock \urlprefix\url{http://doi.org/10.1103/RevModPhys.84.1419}

\bibitem{marzari.vanderbilt.97}
N.~Marzari, D.~Vanderbilt, Phys. Rev. B \textbf{56}, 12847 (1997).
\newblock \doi{10.1103/PhysRevB.56.12847}.
\newblock \urlprefix\url{http://doi.org/10.1103/PhysRevB.56.12847}

\bibitem{souza.marzari.01}
I.~Souza, N.~Marzari, D.~Vanderbilt, Phys. Rev. B \textbf{65}, 035109 (2001).
\newblock \doi{10.1103/PhysRevB.65.035109}.
\newblock \urlprefix\url{https://doi.org/10.1103/PhysRevB.65.035109}

\bibitem{mostofi.yates.08}
A.A. Mostofi, J.R. Yates, Y.S. Lee, I.~Souza, D.~Vanderbilt, N.~Marzari,
  Comput. Phys. Commun. \textbf{178}, 685 (2008).
\newblock \doi{10.1016/j.cpc.2007.11.016}.
\newblock \urlprefix\url{https://doi.org/10.1016/j.cpc.2007.11.016}

\bibitem{mostofi.yates.14}
A.A. Mostofi, J.R. Yates, G.~Pizzi, Y.S. Lee, I.~Souza, D.~Vanderbilt,
  N.~Marzari, Comput. Phys. Commun. \textbf{185}, 2309 (2014).
\newblock \doi{10.1016/j.cpc.2014.05.003}.
\newblock \urlprefix\url{https://doi.org/10.1016/j.cpc.2014.05.003}

\bibitem{pizzi.vitale.20}
G.~Pizzi, V.~Vitale, R.~Arita, S.~Bl\"{u}gel, F.~Freimuth, G.~G{\'{e}}ranton,
  M.~Gibertini, D.~Gresch, C.~Johnson, T.~Koretsune, J.~Iba{\~{n}}ez-Azpiroz,
  H.~Lee, J.M. Lihm, D.~Marchand, A.~Marrazzo, Y.~Mokrousov, J.I. Mustafa,
  Y.~Nohara, Y.~Nomura, L.~Paulatto, S.~Ponc{\'{e}}, T.~Ponweiser, J.~Qiao,
  F.~Th\"{o}le, S.S. Tsirkin, M.~Wierzbowska, N.~Marzari, D.~Vanderbilt,
  I.~Souza, A.A. Mostofi, J.R. Yates, J. Phys.: Condens. Matter \textbf{32},
  165902 (2020).
\newblock \doi{10.1088/1361-648x/ab51ff}.
\newblock \urlprefix\url{https://doi.org/10.1088/1361-648x/ab51ff}

\bibitem{sancho.sancho.85}
M.P.L. Sancho, J.M.L. Sancho, J.M.L. Sancho, J.~Rubio, J. Phys. F: Met. Phys.
  \textbf{15}, 851 (1985).
\newblock \doi{10.1088/0305-4608/15/4/009}.
\newblock \urlprefix\url{https://doi.org/10.1088/0305-4608/15/4/009}

\bibitem{wu.zhang.18}
Q.S. Wu, S.N. Zhang, H.F. Song, M.~Troyer, A.A. Soluyanov, Comput. Phys.
  Commun. \textbf{224}, 405 (2018).
\newblock \doi{10.1016/j.cpc.2017.09.033}.
\newblock \urlprefix\url{https://doi.org/10.1016/j.cpc.2017.09.033}

\bibitem{blochl.94}
P.E. Bl\"ochl, Phys. Rev. B \textbf{50}, 17953 (1994).
\newblock \doi{10.1103/PhysRevB.50.17953}.
\newblock \urlprefix\url{http://doi.org/10.1103/PhysRevB.50.17953}

\bibitem{kresse.hafner.94}
G.~Kresse, J.~Hafner, Phys. Rev. B \textbf{49}, 14251 (1994).
\newblock \doi{10.1103/PhysRevB.49.14251}.
\newblock \urlprefix\url{http://doi.org/10.1103/PhysRevB.49.14251}

\bibitem{kresse.furthmuller.96}
G.~Kresse, J.~Furthm\"uller, Phys. Rev. B \textbf{54}, 11169 (1996).
\newblock \doi{10.1103/PhysRevB.54.11169}.
\newblock \urlprefix\url{http://doi.org/10.1103/PhysRevB.54.11169}

\bibitem{kresse.joubert.99}
G.~Kresse, D.~Joubert, Phys. Rev. B \textbf{59}, 1758 (1999).
\newblock \doi{10.1103/PhysRevB.59.1758}.
\newblock \urlprefix\url{http://doi.org/10.1103/PhysRevB.59.1758}

\bibitem{monkhorst.pack.76}
H.J. Monkhorst, J.D. Pack, Phys. Rev. B \textbf{13}, 5188 (1976).
\newblock \doi{10.1103/PhysRevB.13.5188}.
\newblock \urlprefix\url{http://doi.org/10.1103/PhysRevB.13.5188}

\bibitem{ptok.kapcia.17}
A.~Ptok, K.J. Kapcia, P.~Piekarz, A.M. Ole{\'{s}}, New J. Phys. \textbf{19},
  063039 (2017).
\newblock \doi{10.1088/1367-2630/aa6d9d}.
\newblock \urlprefix\url{https://doi.org/10.1088/1367-2630/aa6d9d}

\bibitem{hirschfeld.korshunov.11}
P.J. Hirschfeld, M.M. Korshunov, I.I. Mazin, Rep. Prog. Phys. \textbf{74},
  124508 (2011).
\newblock \doi{10.1088/0034-4885/74/12/124508}.
\newblock \urlprefix\url{https://doi.org/10.1088/0034-4885/74/12/124508}

\bibitem{ptok.kapcia.20f}
A.~Ptok, K.J. Kapcia, P.~Piekarz, Front. Phys. \textbf{8} (2020).
\newblock \doi{10.3389/fphy.2020.00284}.
\newblock \urlprefix\url{https://doi.org/10.3389/fphy.2020.00284}

\bibitem{ma.wang.22}
X.~Ma, G.~Wang, R.~Liu, T.~Yu, Y.~Peng, P.~Zheng, Z.~Yin, Phys. Rev. B
  \textbf{106}, 115114 (2022).
\newblock \doi{10.1103/PhysRevB.106.115114}.
\newblock \urlprefix\url{https://doi.org/10.1103/PhysRevB.106.115114}

\bibitem{mizuguchi.hara.10}
Y.~Mizuguchi, Y.~Hara, K.~Deguchi, S.~Tsuda, T.~Yamaguchi, K.~Takeda,
  H.~Kotegawa, H.~Tou, Y.~Takano, upercond. Sci. Technol. \textbf{23}, 054013
  (2010).
\newblock \doi{10.1088/0953-2048/23/5/054013}.
\newblock \urlprefix\url{https://doi.org/10.1088/0953-2048/23/5/054013}

\bibitem{kang.birol.17}
C.J. Kang, T.~Birol, G.~Kotliar, Phys. Rev. B \textbf{95}, 014511 (2017).
\newblock \doi{10.1103/PhysRevB.95.014511}.
\newblock \urlprefix\url{https://doi.org/10.1103/PhysRevB.95.014511}

\bibitem{zhang.wang.21}
T.~Zhang, Y.~Wang, J.~Xian, S.~Wang, J.~Fang, S.~Duan, X.~Gao, H.~Song, H.~Liu,
  Matter Radiat. Extremes \textbf{6}, 068401 (2021).
\newblock \doi{10.1063/5.0059360}.
\newblock \urlprefix\url{https://doi.org/10.1063/5.0059360}

\bibitem{pettifor.77}
D.G. Pettifor, J. Phys. F: Met. Phys. \textbf{7}, 613 (1977).
\newblock \doi{10.1088/0305-4608/7/4/013}.
\newblock \urlprefix\url{https://doi.org/10.1088/0305-4608/7/4/013}

\bibitem{moriarty.92}
J.A. Moriarty, Phys. Rev. B \textbf{45}, 2004 (1992).
\newblock \doi{10.1103/PhysRevB.45.2004}.
\newblock \urlprefix\url{https://doi.org/10.1103/PhysRevB.45.2004}

\bibitem{grad.blaha.00}
G.B. Grad, P.~Blaha, J.~Luitz, K.~Schwarz, A.~Fern\'andez~Guillermet, S.J.
  Sferco, Phys. Rev. B \textbf{62}, 12743 (2000).
\newblock \doi{10.1103/PhysRevB.62.12743}.
\newblock \urlprefix\url{https://doi.org/10.1103/PhysRevB.62.12743}

\bibitem{lifshitz.60}
I.M. Lifshitz, Zh. Eksp. Teor. Fiz. \textbf{38}, 1569 (1960).
\newblock [Sov. Phys. JETP \textbf{11}, 1130--1135 (1960)]

\bibitem{norman.lin.10}
M.R. Norman, J.~Lin, A.J. Millis, Phys. Rev. B \textbf{81}, 180513 (2010).
\newblock \doi{10.1103/PhysRevB.81.180513}.
\newblock \urlprefix\url{https://doi.org/10.1103/PhysRevB.81.180513}

\bibitem{lebouf.boironleyraund.11}
D.~LeBoeuf, N.~Doiron-Leyraud, B.~Vignolle, M.~Sutherland, B.J. Ramshaw,
  J.~Levallois, R.~Daou, F.~Lalibert\'e, O.~Cyr-Choini\`ere, J.~Chang, Y.J. Jo,
  L.~Balicas, R.~Liang, D.A. Bonn, W.N. Hardy, C.~Proust, L.~Taillefer, Phys.
  Rev. B \textbf{83}, 054506 (2011).
\newblock \doi{10.1103/PhysRevB.83.054506}.
\newblock \urlprefix\url{https://doi.org/10.1103/PhysRevB.83.054506}

\bibitem{sato.nakayama.09}
T.~Sato, K.~Nakayama, Y.~Sekiba, P.~Richard, Y.M. Xu, S.~Souma, T.~Takahashi,
  G.F. Chen, J.L. Luo, N.L. Wang, H.~Ding, Phys. Rev. Lett. \textbf{103},
  047002 (2009).
\newblock \doi{10.1103/PhysRevLett.103.047002}.
\newblock \urlprefix\url{https://doi.org/10.1103/PhysRevLett.103.047002}

\bibitem{nakayama.sato.11}
K.~Nakayama, T.~Sato, P.~Richard, Y.M. Xu, T.~Kawahara, K.~Umezawa, T.~Qian,
  M.~Neupane, G.F. Chen, H.~Ding, T.~Takahashi, Phys. Rev. B \textbf{83},
  020501 (2011).
\newblock \doi{10.1103/PhysRevB.83.020501}.
\newblock \urlprefix\url{https://doi.org/10.1103/PhysRevB.83.020501}

\bibitem{maleab.shimojima.12}
W.~Malaeb, T.~Shimojima, Y.~Ishida, K.~Okazaki, Y.~Ota, K.~Ohgushi, K.~Kihou,
  T.~Saito, C.H. Lee, S.~Ishida, M.~Nakajima, S.~Uchida, H.~Fukazawa,
  Y.~Kohori, A.~Iyo, H.~Eisaki, C.T. Chen, S.~Watanabe, H.~Ikeda, S.~Shin,
  Phys. Rev. B \textbf{86}, 165117 (2012).
\newblock \doi{10.1103/PhysRevB.86.165117}.
\newblock \urlprefix\url{https://doi.org/10.1103/PhysRevB.86.165117}

\bibitem{xu.richard.13}
N.~Xu, P.~Richard, X.~Shi, A.~van Roekeghem, T.~Qian, E.~Razzoli, E.~Rienks,
  G.F. Chen, E.~Ieki, K.~Nakayama, T.~Sato, T.~Takahashi, M.~Shi, H.~Ding,
  Phys. Rev. B \textbf{88}, 220508 (2013).
\newblock \doi{10.1103/PhysRevB.88.220508}.
\newblock \urlprefix\url{https://doi.org/10.1103/PhysRevB.88.220508}

\bibitem{hodovanets.liu.14}
H.~Hodovanets, Y.~Liu, A.~Jesche, S.~Ran, E.D. Mun, T.A. Lograsso, S.L. Bud'ko,
  P.C. Canfield, Phys. Rev. B \textbf{89}, 224517 (2014).
\newblock \doi{10.1103/PhysRevB.89.224517}.
\newblock \urlprefix\url{https://doi.org/10.1103/PhysRevB.89.224517}

\bibitem{liu.lograsson.14}
Y.~Liu, T.A. Lograsso, Phys. Rev. B \textbf{90}, 224508 (2014).
\newblock \doi{10.1103/PhysRevB.90.224508}.
\newblock \urlprefix\url{https://doi.org/10.1103/PhysRevB.90.224508}

\bibitem{cho.konczykowski.16}
K.~Cho, M.~Ko\'{n}czykowski, S.~Teknowijoyo, M.A. Tanatar, Y.~Liu, T.A.
  Lograsso, W.E. Straszheim, V.~Mishra, S.~Maiti, P.J. Hirschfeld, R.~Prozorov,
  Sci. Adv. \textbf{2}, e1600807 (2016).
\newblock \doi{10.1126/sciadv.1600807}.
\newblock \urlprefix\url{https://doi.org/10.1126/sciadv.1600807}

\bibitem{khan.johnson.14}
S.N. Khan, D.D. Johnson, Phys. Rev. Lett. \textbf{112}, 156401 (2014).
\newblock \doi{10.1103/PhysRevLett.112.156401}.
\newblock \urlprefix\url{https://doi.org/10.1103/PhysRevLett.112.156401}

\bibitem{benhabib.sacuto.15}
S.~Benhabib, A.~Sacuto, M.~Civelli, I.~Paul, M.~Cazayous, Y.~Gallais, M.A.
  M\'easson, R.D. Zhong, J.~Schneeloch, G.D. Gu, D.~Colson, A.~Forget, Phys.
  Rev. Lett. \textbf{114}, 147001 (2015).
\newblock \doi{10.1103/PhysRevLett.114.147001}.
\newblock \urlprefix\url{https://doi.org/10.1103/PhysRevLett.114.147001}

\bibitem{liu.kondo.10}
C.~Liu, T.~Kondo, R.M. Fernandes, A.D. Palczewski, E.D. Mun, N.~Ni, A.N.
  Thaler, A.~Bostwick, E.~Rotenberg, J.~Schmalian, S.L. Bud'ko, P.C. Canfield,
  A.~Kaminski, Nature Phy. \textbf{6}, 419 (2010).
\newblock \doi{10.1038/nphys1656}.
\newblock \urlprefix\url{https://doi.org/10.1038/nphys1656}

\bibitem{harrison.sebastian.07}
N.~Harrison, S.E. Sebastian, C.H. Mielke, A.~Paris, M.J. Gordon, C.A. Swenson,
  D.G. Rickel, M.D. Pacheco, P.F. Ruminer, J.B. Schillig, J.R. Sims, A.H.
  Lacerda, M.T. Suzuki, H.~Harima, T.~Ebihara, Phys. Rev. Lett. \textbf{99},
  056401 (2007).
\newblock \doi{10.1103/PhysRevLett.99.056401}.
\newblock \urlprefix\url{https://doi.org/10.1103/PhysRevLett.99.056401}

\bibitem{purcell.graf.09}
K.M. Purcell, D.~Graf, M.~Kano, J.~Bourg, E.C. Palm, T.~Murphy, R.~McDonald,
  C.H. Mielke, M.M. Altarawneh, C.~Petrovic, R.~Hu, T.~Ebihara, J.~Cooley,
  P.~Schlottmann, S.W. Tozer, Phys. Rev. B \textbf{79}, 214428 (2009).
\newblock \doi{10.1103/PhysRevB.79.214428}.
\newblock \urlprefix\url{https://doi.org/10.1103/PhysRevB.79.214428}

\bibitem{daou.bergemann.06}
R.~Daou, C.~Bergemann, S.R. Julian, Phys. Rev. Lett. \textbf{96}, 026401
  (2006).
\newblock \doi{10.1103/PhysRevLett.96.026401}.
\newblock \urlprefix\url{https://doi.org/10.1103/PhysRevLett.96.026401}

\bibitem{schlottmann.11}
P.~Schlottmann, Phys. Rev. B \textbf{83}, 115133 (2011).
\newblock \doi{10.1103/PhysRevB.83.115133}.
\newblock \urlprefix\url{https://doi.org/10.1103/PhysRevB.83.115133}

\bibitem{ptok.cichy.17}
A.~Ptok, A.~Cichy, K.~Rodr\'{\i}guez, K.J. Kapcia, Phys. Rev. A \textbf{95},
  033613 (2017).
\newblock \doi{10.1103/PhysRevA.95.033613}.
\newblock \urlprefix\url{https://doi.org/10.1103/PhysRevA.95.033613}

\bibitem{ptok.kapcia.17m}
A.~Ptok, K.J. Kapcia, A.~Cichy, A.M. Ole\'{s}, P.~Piekarz, Sci. Rep.
  \textbf{7}, 41979 (2017).
\newblock \doi{10.1038/srep41979}.
\newblock \urlprefix\url{https://doi.org/10.1038/srep41979}

\bibitem{ptok.18}
A.~Ptok, Acta Phys. Pol. A \textbf{135}, 55 (2019).
\newblock \doi{10.12693/APhysPolA.135.55}.
\newblock \urlprefix\url{https://doi.org/10.12693/APhysPolA.135.55}

\bibitem{sen.guo.20}
S.~Sen, G.Y. Guo, Phys. Rev. Mater. \textbf{4}, 104802 (2020).
\newblock \doi{10.1103/PhysRevMaterials.4.104802}.
\newblock \urlprefix\url{https://doi.org/10.1103/PhysRevMaterials.4.104802}

\bibitem{ghosh.sen.21}
A.~Ghosh, S.~Sen, H.~Ghosh, Comput. Mater. Sci. \textbf{186}, 109991 (2021).
\newblock \doi{10.1016/j.commatsci.2020.109991}.
\newblock \urlprefix\url{https://doi.org/10.1016/j.commatsci.2020.109991}

\bibitem{ghosh.ghosh.22}
S.~Ghosh, H.~Ghosh, Comput. Mater. Sci. \textbf{204}, 111170 (2022).
\newblock \doi{10.1016/j.commatsci.2021.111170}.
\newblock \urlprefix\url{https://doi.org/10.1016/j.commatsci.2021.111170}

\bibitem{li.li.12}
W.~Li, J.~Li, J.X. Zhu, Y.~Chen, C.S. Ting, Euro. Phys. Lett. \textbf{99},
  57006 (2012).
\newblock \doi{10.1209/0295-5075/99/57006}.
\newblock \urlprefix\url{https://doi.org/10.1209/0295-5075/99/57006}

\bibitem{momma.izumi.11}
K.~Momma, F.~Izumi, J. Appl. Crystallogr. \textbf{44}, 1272 (2011).
\newblock \doi{10.1107/S0021889811038970}.
\newblock \urlprefix\url{https://doi.org/10.1107/S0021889811038970}

\bibitem{kokalj.99}
A.~Kokalj, J. Mol. Graphics Modelling \textbf{17}, 176 (1999).
\newblock \doi{10.1016/S1093-3263(99)00028-5}.
\newblock \urlprefix\url{http://doi.org/10.1016/S1093-3263(99)00028-5}

\end{thebibliography}

\end{document}